# What's in a Country Name? - Twitter Hashtag Analysis of #singapore

Aravind Sesagiri Raamkumar

Nanyang Technological University

Author Note

The author wishes to specially thank his PhD supervisors Professor Schubert Foo and Assistant Professor Natalie Pang for their helpful insights during the execution of the research study which was conducted for the fulfilment of the Independent Study in Information course.

Correspondence concerning this paper can be sent to Aravind Sesagiri Raamkumar, Wee Kim Wee School of Communication and Information, Nanyang Technological University, 04-39, 31 Nanyang Link, Singapore-637718.

**ABSTRACT**

Twitter as a micro-blogging platform rose to instant fame mainly due to its minimalist features that allow seamless communication between users. As the conversations grew thick and faster, a placeholder feature called as Hashtags became important as it captured the themes behind the tweets. Prior studies have investigated the conversation dynamics, interplay with other media platforms and communication patterns between users for specific event-based hashtags such as the #Occupy movement. Commonplace hashtags which are used on a daily basis have been largely ignored due to their seemingly innocuous presence in tweets and also due to the lack of connection with real-world events. However, it can be postulated that utility of these hashtags is the main reason behind their continued usage. This study is aimed at understanding the rationale behind the usage of a particular type of commonplace hashtags:-location hashtags such as country and city name hashtags. Tweets with the hashtag #singapore were extracted for a week's duration. Manual and automatic tweet classification was performed along with social network analysis, to identify the underlying themes. Seven themes were identified. Findings indicate that the hashtag is prominent in tweets about local events, local news, users' current location and landmark related information sharing. Users who share content from social media sites such as Instagram make use of the hashtag in a more prominent way when compared to users who post textual content. News agencies, commercial bodies and celebrities make use of the hashtag more than common individuals. Overall, the results show the non-conversational nature of the hashtag. The findings are to be validated with other country names and cross-validated with hashtag data from other social media platforms.

# Introduction

Twitter has become one of the most popular Online Social Networks (OSN) in recent times (Zhao & Rosson, 2009). It has popularized the concept of micro-blogging(Java, Song, Finin, & Tseng, 2007) and brought about its relevance in both interpersonal and public communication spheres. Tweets are posts with 140 character limit. Tweets have been the focus of research studies from 2008 (Krishnamurthy, Gill& Arlitt, 2008). People use tweets for multiple purposes such as status updates, conversations with other users, endorsing opinions ('retweet' and 'favourite' options), promotions and even spamming (Benevenuto, Magno, Rodrigues, & Almeida, 2010). Hashtags (words starting with '#' symbol) are used in twitter as a placeholder with multiple purposes. Due to the 140 character limit in tweets, it is important to have an indicator in the tweet to show its representativeness to an idea or concept. Hashtags are prominently used to show the inclusive nature of tweets to a particular topic of conversation. For example, the hashtag #Occupy was used by twitter users during the famous Occupy Movement in 2011[1]. Hashtags can refer to places such as country names (ex: #singapore, #india) which are used on a daily basis. Hashtags are also used as reference to people (ex: #Obama, #stevejobs). Apart from aiding users in conversation, hashtags are used to indicate the main theme of a tweet. For example, the hashtag #review is used in tweets indicating that the tweet text is related to a movie review. Therefore, hashtags are ideal candidates for indexing so as to speed up information retrieval.

Current hashtag studies have taken two approaches, either they have concentrated on event-oriented hashtags such as the ones used in US presidential elections(Lin, Margolin, Keegan, Baronchelli, & Lazer, 2012) or they have used a hashtag agnostic approach where a random extract of twitter data is taken and analysed(Pöschko, 2011). Commonplace hashtags which trend in twitter on a daily basis have not been studied extensively. It would be interesting to study the rationale behind the usage of country/city hashtags and their contribution to conversations at a holistic level. It is observed that some of the popular and frequently used hashtags refer to place names and people names[2]. A place name can be related to a location, town, city or a country. Country names are most often used in this category. These country hashtags are added to show that the particular tweet's content is related to the country. This study takes an explorative approach to understand the dynamics around #singapore by using Text Classification and Social Network Analysis (SNA) based techniques.

---

[1] http://www.technologyreview.com/view/426079/how-occupy-wall-street-occupied-twitter-too/

## Literature Review

**Twitter Research**

Twitter was launched in 2006 as a microblogging platform that facilitated users in sharing and consuming information about day-to-day happenings and opinions on topics. It was a unique product during the time of its introduction due to its character limit on user posts (users are allowed to post tweet messages within 140 characters limit). Twitter has become immensely popular (Rank 10 in Alexa web rankings[3], Rank 3 among social networking sites[4]) which has to lead to regional spinoffs such as Sina Weibo[5]. Noticing the dynamics around the interactions in twitter, academic research in twitter started in 2008-2009(Krishnamurthy et al., 2008). Twitter research has been surveyed and summarized in (Cheong & Lee, 2010; Cheong & Ray, 2011; Williams, Terras, & Warwick, 2013). Research has furthered in different directions with varied focus such as organising information (Sriram, Fuhry, Demir, Ferhatosmanoglu, & Demirbas, 2010), understanding trends and convergence events from a communications perspective(Lin et al., 2012), usage of twitter data in practical applications (ex: governments, activism) (Bruns & Burgess, 2011), cross-application of twitter data (in cross-platform recommendations(Abel, Herder, Houben, & Henze, 2011)) and traditional computer science oriented focus on information retrieval (Magnani, Montesi, Nunziante, & Rossi, 2011) and semantics(Abel, Celik, Houben, & Siehndel, 2011).

---

[2] http://www.trendinalia.com/twitter-trending-topics/singapore/singapore-131126.html
[3] http://www.alexa.com/siteinfo/twitter.com
[4] http://www.socialnetworkingwatch.com/international-social-netw.html
[5] Sina Weibo http://www.weibo.com/

**Tweet Classification**

The two high level entities in twitter are User and Message (Cheong & Lee, 2010). Recent research has introduced two additional entities Technology and Concept (the central topic being addressed in the tweet)(Williams et al., 2013). This classification scheme has been used in studying twitter data. On the topic of tweet[6] classification, past research has identified many categories which differ based on the method of classification, amount of data, period of data and frame of reference. The categories identified by earlier research studies are presented in Table 1 (a&b). Categories such as News, Information sharing, Events, Opinions and Promotions appear to be common across the schemes. The variation in schemes is mainly due to the vocabulary used for naming the categories and the purpose of classification. There has been a lack of consolidation across studies except for the work of Dann(2010) where four earlier classification schemes have been combined to form a new scheme with six generic categories. It is to be noted that all these classification attempts have not used hashtag as the frame of reference.

Table 1 (a)

*Classification Schemes from Previous Twitter Studies(2007-2010)*

| Java et al(2007) | Jansen et al (2009) | Honeycutt & Herring (2009) | Pear Analytics (2009) | Horn (2010) |
|---|---|---|---|---|
| Conversations | Info seeking | About addressee | Mainstream News | C1: News, |
| URL sharing | Info providing | Advertise | Spam | Events, |
| News reporting | Comment/Sentiment | Exhort | Self-promotion of | Company |
| Daily chatter | | Info for others | businesses | C2: Factual, |
| | | Info for self | Babble | Opionated |
| | | Meta-commentary | Conversations | |
| | | Media use | Pass-along messages | |
| | | Express opinion | (retweets) | |
| | | Other's experience | | |
| | | Self experience | | |
| | | Solicit info | | |
| | | Other miscellaneous | | |

---

[6]Tweet is the short post or message posted by the user in Twitter

Table 1 (b)

*Classification Schemes from Previous Twitter Studies (2010-2011)*

| Sriram et al (2010) | Dann (2010) | Sandra et al (2010) | Naaman et al (2010) | Rosa (2011) |
|---|---|---|---|---|
| News | Conversational | Movies | Info sharing | News |
| Opinions | Pass along | Books | Self-promotion | Sports |
| Deals | News | Music | Opinions/Complaints | Science & Technology |
| Events | Status | Apps | Statements & Random Thoughts | Entertainment |
| Private Messages | Phatic | Games | Me now | Money/Business |
|  | Spam |  | Question to followers | Just for Fun |
|  |  |  | Presence Maintenance |  |
|  |  |  | Anecdote |  |

**Hashtag Studies**

Hashtag is a keyword which starts with the symbol '#'. It is mainly used for categorizing content and joining conversations on various topics (Huang, Thornton, & Efthimiadis, 2010). Hashtags serve the same purpose as tags made famous by web 2.0 services such as Flickr[7] and Delicious[8]. Even though, users need not necessarily add hashtags to their tweets, it is generally observed that regular users add hashtags to most of their tweets. Global political events are represented in twitter through hashtags, some of the popular ones include #occupy, #OWS and #Syria. #sghaze was quite popular in Singapore during the period of August 2013[9]. The analysis of behavior around hashtags was done as part of an earlier study on conversational tagging (Huang et al., 2010). The authors use statistical measures such as standard deviation, skewness and kurtosis to study the popularity of hashtags and the scenarios in which hashtags gain traction. Pöschko (2011) performed an exploratory study on hashtags by analysing 29 million tweets which involved tweet classification, studying hashtag co-occurrences, part-of-speech tagging and SNA based clustering thereby highlighting different ways of dissecting the tweet data to gain insights. Yang et al (2012) developed a machine learning model to predict the future adoption of hashtags by users, by combining the two use-cases under which a hashtag is used by users.

---

[7] Flickr http://www.flickr.com/
[8] Delicious https://delicious.com/
[9] Twitter search for #sghaze https://twitter.com/search?q=%23sghaze&src=typd

The two use-cases are content organisation and community participation. Measures such as relevance, preference, prestige and influence were used as the main features for the machine learning model. A similar albeit technically focussed approach by (Tsur & Rappoport, 2012) used more number of features to predict the spread of ideas (hashtag) in twitter environment. Bruns & Burgess (2011) used social network analysis based techniques to study the growth and decline of conversations happening around hashtags at different points of time and raise the need for a detailed catalogue to better understand the patterns of interaction. Lin et al (2012) did a broader study by analysing 256 hashtags related to the US presidential elections for understanding the growth, survival and context of their usage. They put forth a two-way classification of hashtags with the categories 'Winners' and 'Also-rans' and introduced a theoretical framework to understand the adoption behaviour of user-generated content. As seen from earlier studies, the focus has been largely on event-based hashtags. The dynamics around commonplace hashtags are yet to be explored.

## Research Questions

It is apparent from the earlier studies that hashtags play a focal role in directing conversations in twitter. Explorative hashtag studies (Pöschko, 2011)have taken a generalized approach by not looking at a particular type of hashtags. In-depth studies on hashtags so far have been based on political events which are of periodic nature (Bruns & Burgess, 2011; Lin et al., 2012). Therefore, there is a necessity to explore the dynamics around commonplace hashtags that are used on a regular basis. Hashtags which are about a place (location) are quite common trending topics in twitter. Not much is known about the rationale behind their usage. In this study, hashtag with country names will be studied, particularly in the case of #singapore. It is to be noted that in the case of Singapore, the country and city name are the same.

The overarching research question for the current study is *Why do users make use of the hashtag #singapore?* The specific research questions are stated below:-

*RQ1a: What are the categories that represent the tweets and does the classification scheme built using #singapore as a frame of reference defer from the existing classification schemes and why?*

Justification for RQ1a: The tweet classification approaches have their learning mechanism based on the whole tweet content. It would be interesting to see if the classification performed by keeping the hashtag as the frame of reference, differs from the existing schemes.

*RQ1b: What are the relationships between #singapore and other co-occurring hashtags?*

*RQ1c: Does the provenance data of the tweets provide any new insights?*

*RQ1d: What is the communication pattern between users using #singapore?*

## Methods

**Data Collection**

Twitter data extraction service *TweetArchivist*[10] was used to extract data for the hashtag '#singapore' for the period between August 26th and September 1st 2013. Table 2 provides the tweet count for #singapore for the dates in the selected period along with the other dates from the complete data extraction period provided by the extraction service. It was observed that the tweet contribution was highest on Fridays and weekends. The sample set originally comprised of 20757 tweets for the 7 day period, of which 17798 were considered for the study as the other remaining tweets were not in English.

---

[10] TweetArchivist http://www.tweetarchivist.com/

Table 2

*Tweet Count Statistics for #singapore*

| Date | Tweet Count | Day of week |
|---|---|---|
| Aug 20 | 1360 | Tue |
| Aug 21 | 2305 | Wed |
| Aug 22 | 2776 | Thur |
| Aug 23 | 3780 | Fri |
| Aug 24 | 2753 | Sat |
| Aug 25 | 2524 | Sun |
| *Aug 26* | *2387* | *Mon* |
| *Aug 27* | *3045* | *Tue* |
| *Aug 28* | *2635* | *Wed* |
| *Aug 29* | *3298* | *Thur* |
| *Aug 30* | *3574* | *Fri* |
| *Aug 31* | *2930* | *Sat* |
| *Sep 1* | *2888* | *Sun* |
| Sep 2 | 3009 | Mon |
| Sep 3 | 2644 | Tue |
| Sep 4 | 2676 | Wed |
| Sep 5 | 2554 | Thur |

**Research Methods and Evaluation**

The exploratory nature of this study demanded the employment of content and structural analysis methods to gain deeper understanding of the data. The three main methods used in the study are text classification (Sebastiani, 2002), social network analysis (Wasserman, 1994) and content analysis (Neuendorf, 2002). Text classification was conducted in two phases. The first phase included manual classification of about 500 tweets by three coders. The categories identified by the first coder, were used by the other two coders. Categories were identified by keeping the keyword '#singapore' as the frame of reference. Inter-coder reliability testing using the kappa coefficient was employed to gauge the agreement between the three coders. Automatic text classification methods were employed in the second phase in order to classify the remaining tweets from the sample set. Maximum Entropy (Nigam, Lafferty, & McCallum, 1999) and Support Vector Machines (SVM) (Joachims, 1998) were the two shortlisted algorithms for the text classification as they displayed high accuracy, precision and recall based on a comparison of different machine learning algorithms conducted before the actual classification phase. Seven features were used for the classification process – *Username, Tweet Text, Tweet Source, User Location, Tweet Hashtags, Tweet Urls and Tweet Url Mentions*. The manually coded tweets from the first coder were used as training set. The statistical programming language R along with the machine learning library RTextTools[11] was used for the automatic text classification. Content analysis was employed on tweets from the manually coded set. Social Network Analysis (SNA) techniques were used to analyse the hashtags and user-mentions[12] data extracted from the tweets. A directed graph for user-mentions data and an undirected graph for hashtags were built using the visualization tool Gephi[13]. Modularity based community detection algorithm (Newman, 2006) was run to identify the underlying communities in the data. In-degree property was used to filter sparse data in the graphs. The node sizes for the graphs were based on Betweeness Centrality(Wasserman, 1994).

## Results

**Manual Tweet Classification**

Seven composite categories were identified after the manual classification process. The first coder identified 23 subcategories which were used for coding by the other two coders. The aggregation of the 23 subcategories to the seven categories was performed at a later stage to reduce sparsity in the assignment of categories to the tweets and facilitate stronger agreement between the coders. The seven categories are *Local Events and News (LEN), Current Location and Landmarks (CLL),Asia Related and Unrelated topics (ARU), Commercial Deal (CD), Tourism and Travel Related (TTR), National Identity and Group Reference (NGR) and Personal Events and Rants (PER)*.Figure1provides the count for the category assignments by the three coders. The most differences were seen in the categories NGR and TTR with the third coder tending to assign more tweets to these categories. Table 3provides data about the inter-coder agreement between the three coders. The kappa value for 3 coders was 0.47 which translates to moderate level of agreement (Gwet, 2012) between the coders. The agreement was not high due to the presence of multiple themes in a single tweet which meant the coders had to assign the tweet to the category that they felt to be appropriate. Lower number of categories at the start of the coding would have also facilitated better agreement.

---

[11] RTextTools http://www.rtexttools.com/
[12] In tweets, users can tag other users so as to meaningfully direct the tweet's content. This feature is called as 'User-mentions'
[13] Gephi https://gephi.org/

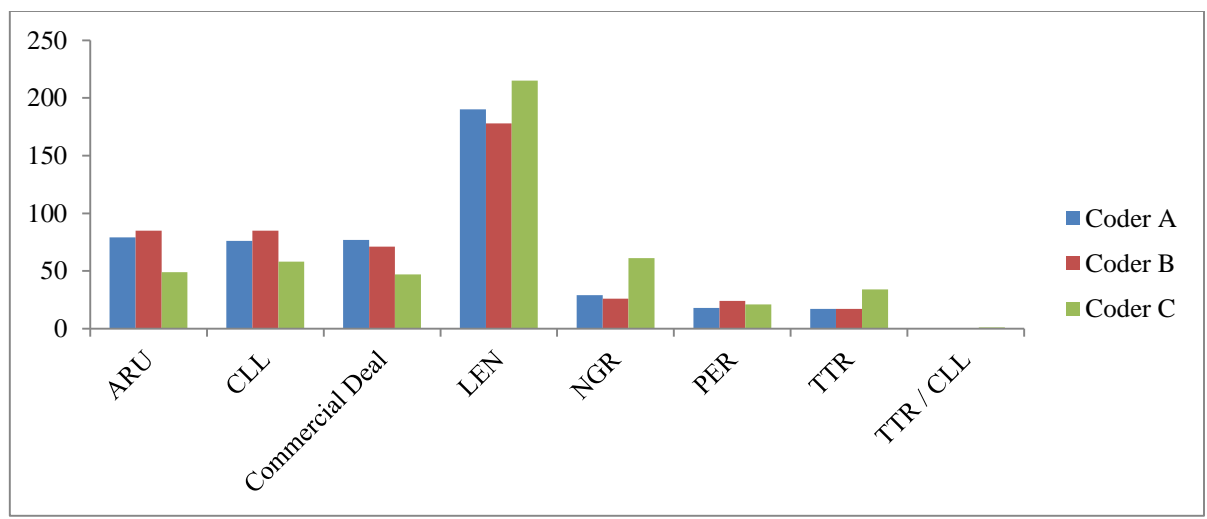

*Figure 1*. Manual Classification – Inter-Coder Agreement

Table 3

*Inter-Coder Agreement Stats*

| Coders | Kappa | Z |
|---|---|---|
| A,B,C | 0.47 | 37.00 |
| A,B | 0.46 | 20.60 |
| B,C | 0.44 | 21.00 |
| A,C | 0.51 | 23.80 |

**Automatic Tweet Classification**

The precision, recall and F-measure values for the machine learning algorithms Maximum Entropy (MAXENT) and Support Vector Machines (SVM) used in training is provided in Table A1 in Appendix. MAXENT was chosen as the final algorithm due to better accuracy rates verified with 10 fold cross validation. Table 4provides the count for the predicted categories for the complete sample set. The categories Local Events and News (LEN) and Current Location and Landmark (CLL) take up majority of the tweets with 28.88% and 27.94% respectively. The Personal Events and Rants (PER) category was the lowest predicted category with 1.15% of allocation.

Table 4

*Tweet Categories Assignment after Automatic Classification*

| Category | Count of Category | Percentage of Category | Retweets |
| --- | --- | --- | --- |
| Local Events and News (LEN) | 5140 | 28.88% | 1755 (34.14%) |
| Current Location and Landmarks (CLL) | 4973 | 27.94% | 1537 (30.90%) |
| Asia Related and Unrelated topics (ARU) | 2822 | 15.86% | 787 (27.88%) |
| Commercial Deals (CD) | 2142 | 12.04% | 280 (13.07%) |
| Tourism and Travel Related (TTR ) | 1302 | 7.32% | 619 (47.54%) |
| National Identity and Group Reference (NGR) | 1214 | 6.82% | 596 (49.09%) |
| Personal Events and Rants (PER) | 205 | 1.15% | 71 (34.63%) |
| *Total* | *17798* | | |

**Social Network Analysis**

Statistics of the two graphs is given in Table A2 (Appendix). The original graphs were filtered based on the in-degree property to prevent sparsity and facilitate interpretation.FigureA1 (Appendix)is the graph constructed using the hashtags data extracted from the tweets. Six communities were detected. One community is related to other Asian countries (brown). Two communities are related to food (violet and dark green). The other three communities are of similar nature and mainly related to locations and Instagram based tweets (red, blue and light green). FigureA2 (Appendix) is the directed graph constructed using the user-mentions data. Majority of the clusters have their focal nodes as news agencies (strait times, channels new asia), celebrities (Pitbull) and commercial group accounts (Marina Bay Sands) which shows that the topic of discussion is mainly centred on content provided by these user accounts with high number of followers.

**Additional Statistics**

FigureA3 (Appendix) is a line graph depicting the hourly tweet count. The activity is high at the start of the business hours (8-9 AM) and it reaches its peak after office hours (5-9 PM). These findings are largely similar to general twitter traffic where the bulk of the postings happen in the evening. FigureA4 (Appendix) is a graph correlating the tweet counts with twitter trends ranking. For the period of August 20 to September 5 2013, it is observed whether the tweet count of either singapore or #singapore has an impact on the ranking of #singapore. The figure shows that #singapore finds its place in the top 20 twitter trends on a consistent basis with a few days going above the top 30. There is no perceivable relationship between the tweet count and the tweet rank i.e. the rank does not improve due to increased number of tweets. To further understand the causal factors behind the ranking of #singapore, the top 20 twitter trend keywords along with the corresponding tweet counts need to be extracted and correlated.

## Discussion

*RQ1a: What are the categories that represent the tweets and does the tweet classification scheme built using #singapore as a frame of reference defer from the existing classification schemes and why?*

*Local Events and News (LEN)*

This theme with a tweet count of 5140 (28.88%) corresponds to the tweets that are about local events and news that are mainly posted by news agencies and commercial bodies. It is quite evident that news agencies use the hashtag more than any other type of user (user accounts personaSingapore - 509 tweets, sgbroadcast- 309 tweets). There are two reasons for this behaviour, the first reason is to gain attention of the public by having a hashtag which is easily relatable and secondly, the hashtag is added to indicate that the tweet content is to be interpreted within the context of Singapore. The local news category subsumes news about sports, weather, entertainment and business. The count of these tweets is higher than any other category due to the regular tweeting done by commercial bodies and also due to the heavy retweeting by normal users (1755 retweets out of 5140). The news subcategory is prevalent in most of the previous studies (Horn, 2010; Java et al., 2007; Rosa et al., 2011; Sriram et al., 2010) while events subcategory in seen in (Horn, 2010).

*Current Location and Landmark (CLL)*

This theme with a tweet count of 4973 (27.94%) corresponds to the tweets that are about the current location of the user and references to landmarks in the locality. The tweets are mainly visual content shared from social sharing sites such as Instagram (2488 out of 4973 tweets). In fact, the finding that majority of the tweets in the sample set were posted from Instagram (refer Table A3 in Appendix) indicates the popularity of the hashtag among users who frequently take photographs in and around Singapore. This category corresponds to the 'Me Now' category from (Naaman et al., 2010). The absence of this category from most of the previous classifications is due to its unique association with a location based hashtag such as #singapore.

*Asia Related and Unrelated topics (ARU)*

This theme with a tweet count of 2822 (15.86%) corresponds to the tweets that are about topics related to other Asian countries and in some cases, topics that do not have relation to #singapore. The spam subcategory plays a major part with just a single spammer contributing to 13% of the total tweets. It is seen that certain Asia-pacific user accounts (Alert_from_Asia, Cherascity)have the tendency to add the hashtag in many tweets regardless of the topic. This theme also covers the tweets that are posted by local news agencies when the topic is related to Asia. The hashtag co-occurrence graph diagram has a specific community (highlighted in brown in Fig A2.1) with #singapore appearing with other Asian country names. These tweets are mostly spams with no value being added by #singapore.

This composite category is not present in previous classification due to its territorial nature, however spam subcategory has already been identified in (Dann, 2010; Pear-Analytics, 2009)

*Commercial Deals (CD)*

This theme with a tweet count of 2142 (12.04%) corresponds to tweets that are posted by commercial bodies based in Singapore, with the intention of marketing and promoting their products to their twitter followers. The activity can be seen as an alternative/compliment to the RSS[14] based push services provided by online portals. Twitter is used by these bodies to push latest information to the users. From the sample set, book stores (sgbookstore – 18% and singaporebook – 9.4%) and job portals (StanChartJobs – 4.5%) use twitter to push new offerings to the public. Similar themed categories are found in most of the earlier studies (Honeycutt & Herring, 2009; Jansen et al., 2009; Naaman et al., 2010; Pear-Analytics, 2009; Sriram et al., 2010). The usage of #singapore in these tweets, is of redundant nature as it can be assumed that the tweets are mostly related to the singapore context. One of the main reasons that commercial bodies persist in continual usage of commonplace hashtags is to leave a digital imprint and capture mind share of users.

*Tourism and Travel Related (TTR)*

This theme with a tweet count of 1302 (7.32%) corresponds to tweets that are related to tourism and travel related information sharing by users. It is closely related to the current location and landmark theme. However, the difference is notable with users posting tweets indicating their travel in and out of Singapore or posting tweets to promote tourism for a particular locality in Singapore. The usage of #singapore in the context of these tweets is very specific as it is directly related to the main topic of the tweet and the presence of #singapore as the first hashtag in 71% of the tweets provides suitable evidence for this claim. This category does not have peers in the classifications of previous studies due to its specific nature.

---

[14]RSS http://www.rssboard.org/rss-specification#whatIsRss

*National Identity and Group References (NGR)*

This theme with a tweet count of 1214 (6.82%) corresponds to tweets that are posted by users as references to fellow Singaporeans and also to convey information or opinion related to the image of Singapore. These are the only set of tweets that are directly addressed to the main topic 'Singapore' from either a geographic or geopolitical viewpoint. This category is dominated with normal user accounts unlike other categories which are mainly represented by group accounts. Greeting messages (ex: good morning wishes) and socio-cultural messages (ex: "Majority of Singaporeans want slower pace of life...") are the types of tweets represented by this category. Similar categories from previous studies are conversational (Dann, 2010), Meta-commentary(Honeycutt & Herring, 2009)and Statements (Naaman et al., 2010).

*Personal Events and Rants (PER)*

This theme with a tweet count of 205 (1.15%) corresponds to tweets that are entirely user specific, referring to a personal event or a personal rant (opinion) about an entity. Frustrations about traffic or a personal communication between smaller groups of people are the candidates for this theme. This category has the lowest allocation amongst all the categories which indicates the lack of its usage by users. However, the tweets from this category are the ones that are extensively studied due to their subjective, user-specific content. A very high percentage of these tweets (48.78%) are sent from the app 'Twitter for IPhone' which underlines the usage by individual users.

The seven categories identified during the classification exercise have both similarities and differences with the categories from previous studies. The common categories include News, Current Location, Commercial Deals, Spams and Group References which shows the pervasive nature of these categories. The novel categories Asia Related, Personal Events, Tourism and National Identity have been newly identified mainly due to the specific nature of a location hashtag such as #singapore and usage of hashtag centric frame of reference. These findings need to be verified by performing a similar exercise with other country and city names. It can be claimed that an all-encompassing classification method needs to have sub-categories to capture the theme of tweets or the classification has to be set at an abstract level.

*RQ1b: What are the relationships between #singapore and other co-occurring hashtags?*

As a precursor towards understanding the relationship between #singapore and other co-occurring hashtags, the importance of #singapore as an independent entity needs to be studied. It is to be noted that there is no restriction on the number of hashtags that can be added to a tweet baring the 140 character limit. Tweets in the sample set have #singapore as the first hashtag in 8087 tweets which constitutes to about 45% of the total tweets in the sample set. The count of tweets containing #singapore by the hashtag position is provided in FigureA5 (Appendix) where the chart data follows a power law distribution. These stats indicate that #singapore plays a primary role in majority of the tweets, at least based on the positioning of hashtags. SNA was used to understand the relationship between #singapore and other hashtags. The undirected graph (FigureA1 in Appendix) constructed using hashtags extracted from the tweets shows the presence of six interrelated communities. There are two communities that are related to food (purple and dark green), corresponding to the CLL theme. Three communities are mainly related to locations and Instagram related tweets (red, blue and light green) corresponding to the themes PER, LEN and CLL. One community is mainly about country names corresponding to ARU theme. The findings from the graph largely corroborates with themes identified during the text classification. It can be claimed that #singapore plays a complimentary and meaningful role in the context of co-occurring hashtags in the tweets. Since most of the categories were identified using the graph from FigureA1 (Appendix), individual category graphs were not required.

*RQ1c: Does the provenance data of tweets provide any new insights?*

Instagram is the most used source in tweets that contain #singapore (refer Table A3 in Appendix) which directly translates to the high number of tweets in the CLL category. This shows the popularity of Instagram as a media sharing platform and also the intent of users to promote their content through twitter. The other major sources are web and smart phones. It is to be noted that twitter was launched as a micro blogging platform for people to share personal opinions and communicate with others in an easier way. Table A4 (Appendix) shows that about 70% of tweets from the sample set contain URLs. This goes to show that people use twitter in the same manner as sharing content through other social media sites such as Tumblr[15] and StumbleUpon[16]. URL sharing behaviour is to signify that *"I as a user have gone through this website and I feel this will be worth reading for you too"*. An earlier study by (Liu, 2013) vindicates this finding. The major presence of URLs in the tweets is due to the content shared from Instagram where the link points to the image or video posted in Instagram. It is claimed that the findings from provenance data is generalizable to other studies involving location hashtags and not for event related hashtags.

*RQ1d: What is the communication pattern between users using #singapore?*

In tweets, users can tag other users so as to meaningfully direct the tweet's content. This feature is called as 'User-mentions'. User-mentions can be noticed whenever a user replies to another user's tweets and also when a user retweets another user's content. FigureA2 (Appendix) shows the presence of four major clusters of users with Pitbull (indegree=749), marinabaysands (indegree=414), STcommunities (indegree=308) and STcom (indegree=295) as the central nodes. The visualization underlines the prevalence of retweeting as major user activity in the sample set as the aforementioned accounts are heavily retweeted by normal users (refer Table A5 in Appendix). This finding can be generalized to a bigger population as earlier findings in the current study have already established the commercial interest behind the usage of #singapore unlike event based hashtags such as #sghaze and #occupy which are mainly used for conversational purposes. The tweets in the sample set indicate very minimal personal communication using #singapore. Average path length of 3.5 indicates longer distances between nodes which translate to communication being restricted within small groups.

---

[15] Tumblr https://www.tumblr.com/
[16] StumbleUpon http://www.stumbleupon.com/

**Limitations**

The results are based on an in-depth analysis of tweets collected for a week's duration which is considered appropriate for the current study's scope. Future studies are planned to be conducted with tweets collected for a longer duration so that generalization is not an issue. In the technical front, the features used in the automatic tweet classification exercise are deemed basic as they are direct fields from the Twitter extract. Complex features are to be devised in future studies to improve the accuracy of the classification algorithms.

## Conclusion and Future Work

The objective of the current study was to identify the rationale behind the usage of the hashtag #singapore in tweets. Seven themes/categories were identified as a part of a tweet classification exercise. The hashtag is prominent in tweets about local events, local news, users' current location and landmark related information sharing. Users who share content from social media sites such as Instagram make use of the hashtag in a more prominent way when compared to users who post textual content. News agencies, commercial bodies and celebrities make use of the hashtag more than common individuals. Similarities and differences with existing tweet classifications were identified along with the justifications for the novel categories. The case for using hashtag as the frame of reference for classification purposes has also been raised. Owing to the relatively small size of the country, the hashtag continues to be one of the top trends on a regular basis due to commercial elements in twitter. SNA based techniques were conducted to further supplement the findings from the classification exercise.

The current study's results are to be validated with similar exercise with different country and city names as the dynamics related to Singapore might not be applicable to other cities and countries. Cross-media validation is to be performed by extracting similar data from platforms such as Google Plus and Facebook as the hashtag has become a common feature across most social media platforms. It would be interesting to study whether users make use of commonplace hashtags with similar intents across platforms.

# Appendix

Table A1

*Performance Metrics of the Machine Learning Algorithms with the Training Data*

| Category | SVM | | | MAXENTROPY | | |
|---|---|---|---|---|---|---|
| | Precision | Recall | F-score | Precision | Recall | F-score |
| Commercial Deal | 0.93 | 0.79 | 0.85 | 0.89 | 0.76 | 0.82 |
| CLL | 0.78 | 0.49 | 0.6 | 0.71 | 0.41 | 0.52 |
| NGR | 0.75 | 0.03 | 0.06 | 0.82 | 0.1 | 0.18 |
| TTR | 0 | 0 | NaN | 0 | 0 | NaN |
| LEN | 0.31 | 0.87 | 0.46 | 0.36 | 0.7 | 0.48 |
| ARU | 0.3 | 0.79 | 0.43 | 0.2 | 0.93 | 0.33 |
| PER | NaN | 0 | NaN | 0.25 | 0.57 | 0.35 |

Table A2

*Graph Statistics*

| Property | Hashtags graph | User-mentions graph |
|---|---|---|
| Type | Undirected | Directed |
| Nodes | 7121 | 8944 |
| Edges | 14299 | 7442 |
| Average Degree | 4.016 | 1.436 |
| Average Path Length | 1.732 | 3.507 |
| Filtered Nodes | 122 | 211 |
| Filtered Edges | 994 | 332 |
| Minimum Degree Filter | 24 | 2 |

*Figure A1*. Undirected Graph of #singapore and Other Co-occurring Hashtags

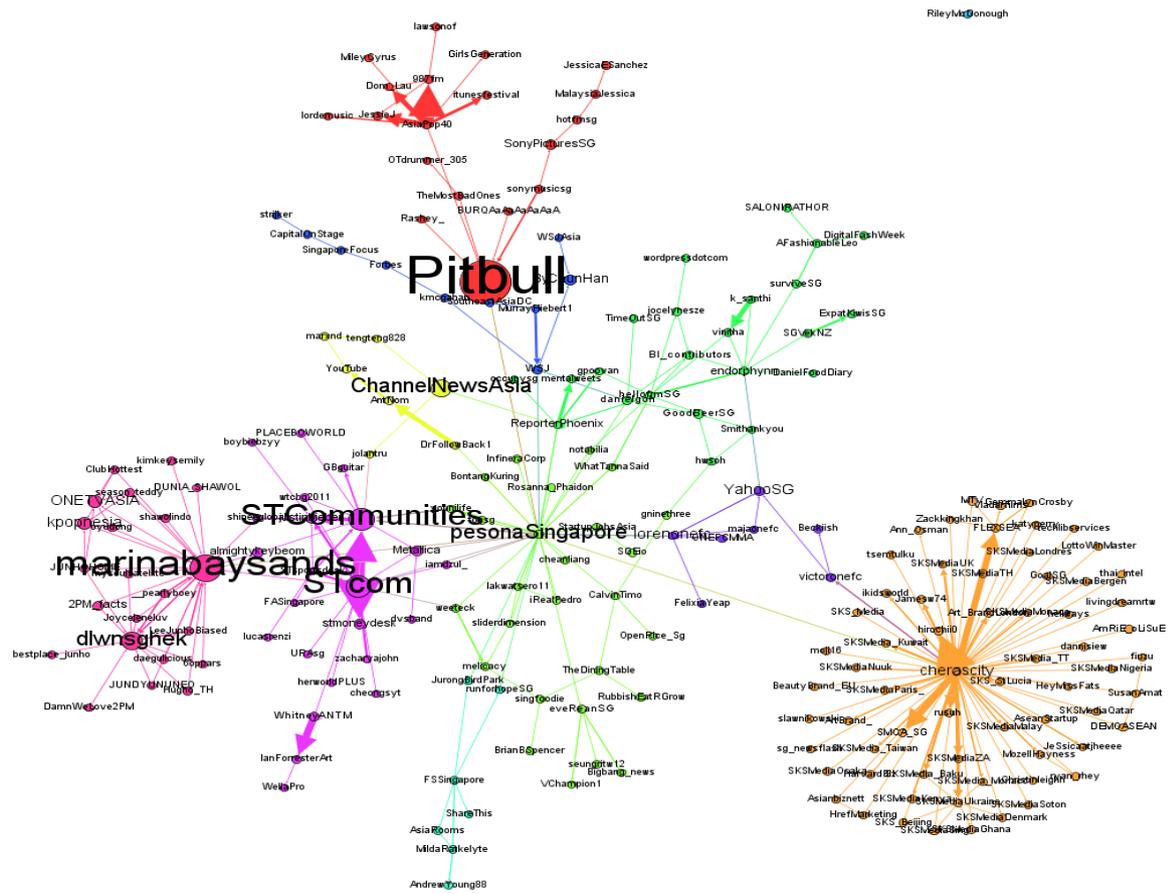

*Figure A2.* Directed Graph Built with 'User-mentions' Data

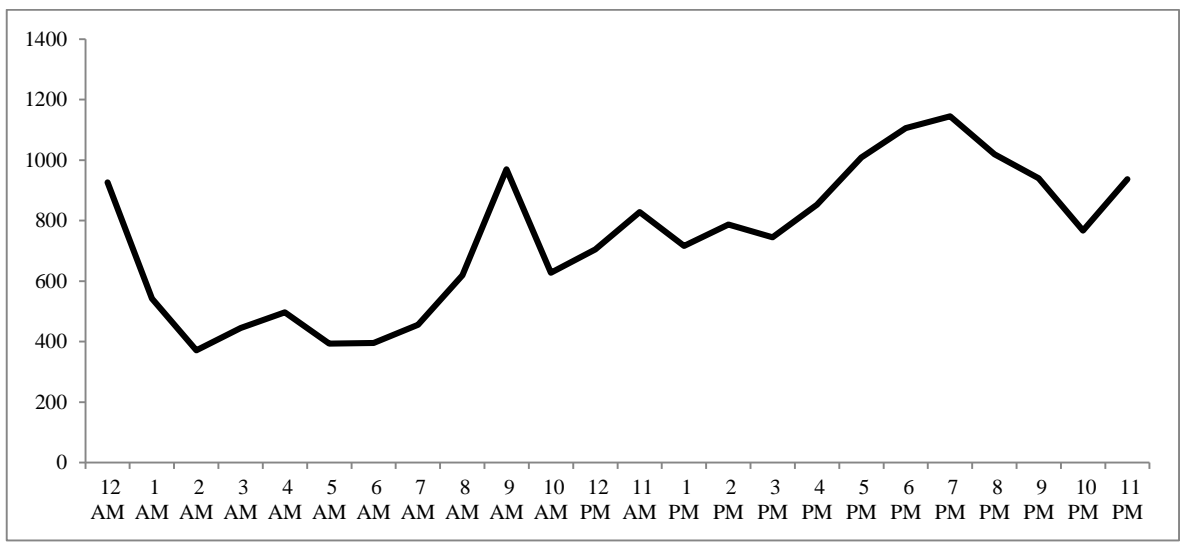

*Figure A3.* Hourly Tweet Count for #singapore

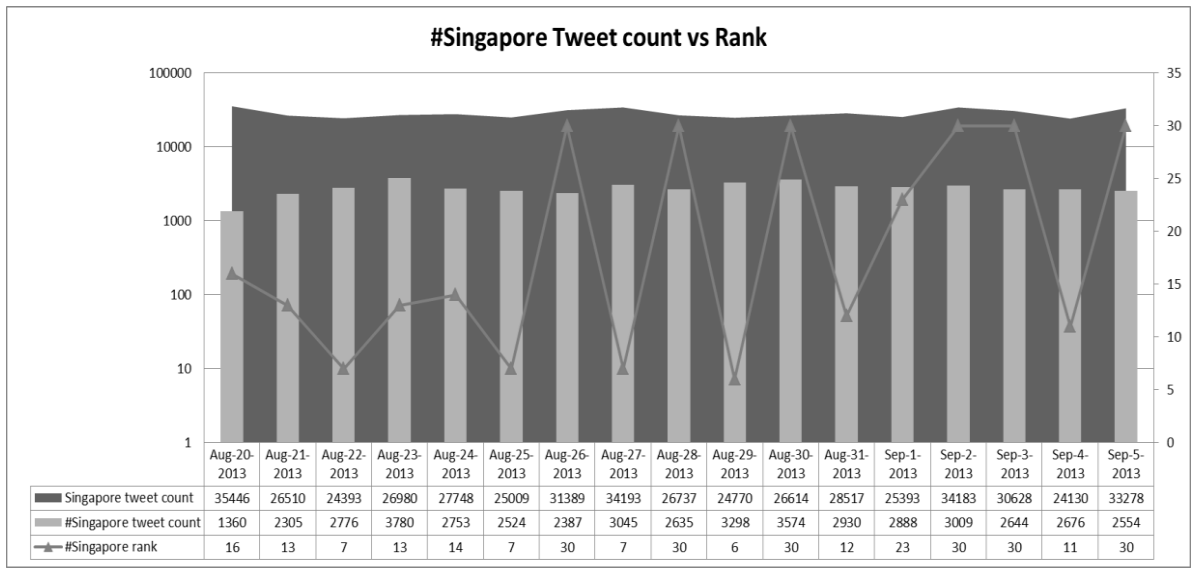

*Figure A4.*Comparisonof Singapore and #singapore Tweet Count with #singapore Trend Tank. Left side X-axis is based on log scale. Tweets counts are based on left side X-axis scale while rank is based on the right side X-axis scale

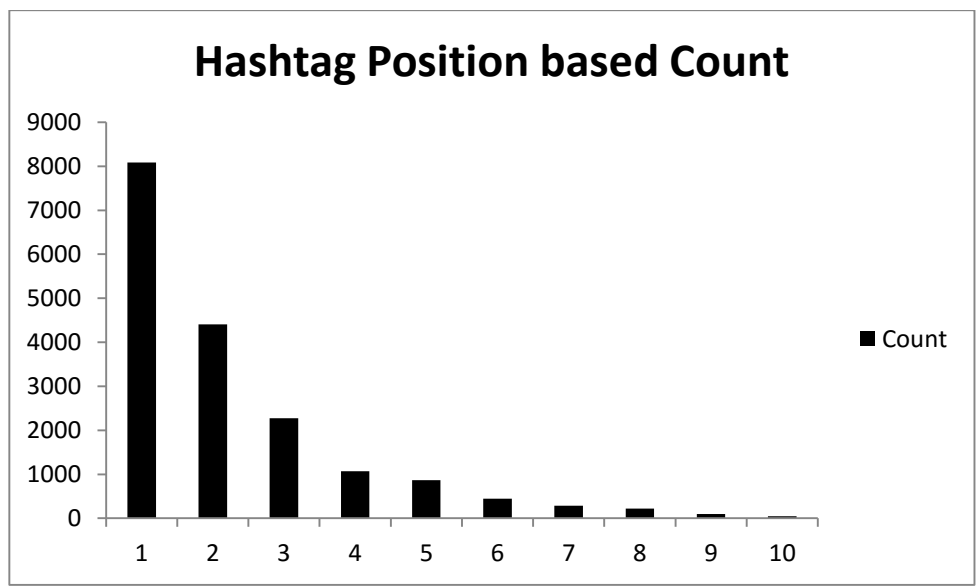

*Figure A5.* Hashtag Position based Count. The bar chart indicates a Power Law Distribution

Table A3

*Top 10 Sources*

| Source | Tweet Count | Percentage |
|---|---|---|
| Instagram | 3206 | 18.01% |
| Web | 1844 | 10.36% |
| Twitter for iPhone | 1629 | 9.15% |
| Twitter for Android | 1432 | 8.05% |
| dlvr.it | 1212 | 6.81% |
| Twitterfeed | 891 | 5.01% |
| RoundTeam | 686 | 3.85% |
| Tweet Old Post | 598 | 3.36% |
| HootSuite | 555 | 3.12% |
| TweetAdder v4 | 547 | 3.07% |

Table A4

*URL Presence in Tweets*

| URL Presence | Tweet Count | Percentage |
|---|---|---|
| Not Present | 5256 | 29.53% |
| Present | 12542 | 70.47% |
| *Total* | *17798* | |

Table A5

*Retweets among the Total Tweets*

| Retweet | Count of Retweet | Percentage |
|---|---|---|
| Normal Tweet | 12165 | 68.35% |
| Retweet | 5633 | 31.65% |
| *Total* | *17798* | |